\documentclass[epjCONF, onecolumn]{svjour}
\usepackage{graphics}
\usepackage[varg]{txfonts} 
\usepackage[latin1]{inputenc}
\session-title{Conference Title, to be filled}
\begin{document}
\title{First Results for the Solar Neighborhood of the Asiago Red Clump Survey}
\author{M. Valentini \inst{1}\fnmsep\thanks{\email{valentini@astro.ulg.ac.be}} \and U. Munari \inst{2} \and T. Saguner \inst{2}  \and K. Freeman \inst{3} \and S. Pasetto \inst{4,5} \and J. Montalb\'{a}n \inst{1} \and E.K. Grebel  \inst{5}}
\institute{ Institute d'Astrophysique et de G\'{e}ophysique, Universit\'{e} de Li\'{e}ge, B-4000 Li\'ege, Belgium
\and INAF-OAPD-Osservatorio Astronomico di Padova, via dell'Osservatorio 8, I-36012 Asiago, Italy \and Mount Stromlo Observatory, Australian National University, Weston Creek, ACT 2611, Australia \and University College London, Dept. of Space \& Climate Physics, Mullard Space Science Lab., UK \and Astronomisches Recheninstitut, Universitaet Heidelberg, Heidelberg, Germany}
\abstract{
The Asiago Red Clump Spectroscopic Survey (ARCS) is an ongoing survey that provides atmospheric parameters,  distances and space velocities of a well selected sample of Red Clump stars distributed along the celestial equator. 
We used the ARCS catalog for a preliminary investigation of the Galactic disk in the Solar Neighborhood, in particular we focused on detection and characterization of moving groups.}
\maketitle
\begin{figure}
\centering
\resizebox{0.57\columnwidth}{!}{
\includegraphics{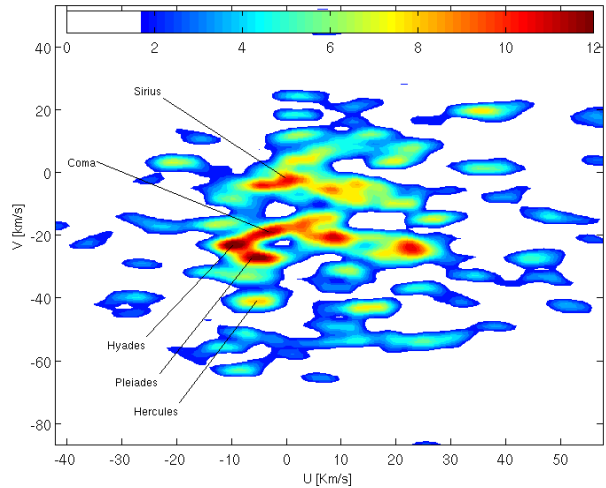} }
\caption{2-D wavelet transform with a Mexican-hat kernel of the U-V distribution of ARCS Red Clump stars. The scale parameter of the analyzing wavelet is 5 km s$^{-1}$, the color scale is adjusted in order to emphasize overdensities.}
\label{fig:1}       
\end{figure}

%
Red Clump stars (hereafter RC stars) are a remarkable and well known tool to investigate the structure and the kinematics of various Galactic subsystems.
First two data releases by Asiago Red Clump surveys (at high  \cite{Val} and intermediate \cite{Sag} resolving power) provide  atmospheric parameters(T$_{eff}$, log(g), [M/H]), spectrophotometric distances and radial velocities for a well selected sample of 439  RC stars, mainly located in a torus extending from 200 to 500 pc from the Sun (future ARCS data releases will explore both inward and outward). Table \ref{tab:1} shows the accuracy of the data provided by the survey. \\
We implemented the catalog with ages computed using the code PARAM, developed by L. Girardi \cite{daS}. The uncertainty on age is up to 40\%-80\%, due to the nature of the RC: isochrones of very different age and metallicity lie closely together in this region. For this reason we used PARAM ages only for a first statistical investigation and we considered only stars with an accuracy on age better than 60\%. The age determination for Red Giants (and hence RC stars) will be improved by recent and exciting asteroseismology findings with data provided by CoRoT and Kepler space missions (\cite{Mos}, \cite{Mont}, \cite{Mig}).\\
A typical analysis of the distribution of ARCS stars in the UV velocity space permits the study of the disk kinematics in the Solar Neighborhood . Different local irregularities that comes out as concentrations of stars in the U-V distribution are the moving groups or stellar streams \cite{Sku}, \cite{Zhao}. Several hypotheses are claimed to explain the origin of moving groups: dispersal stellar clusters, accretion or a resonant mechanism. A homogeneity in chemical abundances and age is expected if the moving group is the debris of star forming aggregates or of an infalling object . On the other hand, an absence of such homogeneities argues for a resonant origin (related to the Galactic bar or spiral arms). \\
A 2-D wavelet transform technique is used to identify moving groups in the U-V plane, as performed by Skuljan \cite{Sku}. We used a Mexican-hat-shaped kernel function with a scale parameter of 5 km s$^{-1}$ (see Figure \ref{fig:1}). Most known moving groups are clearly visible in the ARCS sample. The high accuracy of the data provided by ARCS exclude that detected over-densities are artifacts. In Table \ref{tab:1} we summarized the characteristics of the most prominent moving groups and we compared our results with those of Zhao \cite{Zhao}. There is good agreement between the two works and discrepancies between (U, V) coordinates of the moving groups are mainly due to a different U$_\odot$V$_\odot$W$_\odot$ adopted. The large dispersion in metallicity and age argues in favour of a resonant origin of the moving groups of Sirius-UMa, Coma, Hyades, Pleiades and Hercules. In addiction to these moving groups, Fig. \ref{fig:1} shows other overdensities that are discussed in detail elsewhere (Valentini et al. 2011, submitted).
\begin{table}
\centering
\caption{Accuracy of the ARCS high and medium resolution surveys.}
\label{tab:1}       
\begin{tabular}{ l l l l l }
\hline\noalign{\smallskip}
     & V$_{rad}$ & T$_{eff}$& log(g) & [M/H] \\ 
\noalign{\smallskip}\hline\noalign{\smallskip}
ARCS high &0.5 km s$^{-1}$ & 55 K & 0.12 dex & 0.11 dex \\
 ARCS medium & 1.3 km s$^{-1}$ & 88 K & 0.38 dex & 0.17 dex \\
\noalign{\smallskip}\hline
\end{tabular}
\end{table}
\begin{table}
\centering
\caption{Characteristic U and V velocities, metallicity and age of most prominent features detected in the U-V distribution of ARCS Red Clump stars. Results obtained with ARCS catalog are compared with those present in the catalog of moving groups of  Zhao \cite{Zhao}. }
\label{tab:2}       
\begin{tabular}{ l l l l l l}
\hline\noalign{\smallskip}
ID & \multicolumn{3}{l}{ARCS} & \multicolumn{2}{l}{Zhao et al. (2009)} \\
     & (U,V) & [M/H] & Age            & (U,V) & [M/H] \\
     & km s$^{-1}$ & dex & Gyr & km s$^{-1}$ & dex \\ 
\noalign{\smallskip}\hline\noalign{\smallskip}
Sirius-UMa & ($+$0,$+$7)      & $-$0.11 $\sigma$=0.22  & 1.8 $\sigma$= 3.0 & ($+$10,$-$14) & $-$0.21 $\sigma$=0.15 \\
Coma         & ($-$11,$-$20)    & $-$0.12 $\sigma$=0.20  & 1.5 $\sigma$= 3.0 & ($-$38,$-$17) & $-$0.09 $\sigma$=0.17  \\
Pleiades    & ($-$8,$-$22)      & $-$0.18 $\sigma$=0.26  & 3.0 $\sigma$= 1.8 & ($-$15,$-$23) & $-$0.17 $\sigma$=0.17  \\
Hercules    & ($-$10,$-$40)   & $-$0.10 $\sigma$=0.29  & 5.0 $\sigma$= 3.0 & ($-$35,$-$51) & $-$0.16 $\sigma$=0.20  \\
\noalign{\smallskip}\hline
\end{tabular}
\end{table}

\end{document}